# Generation of Low Divergent High Power Supercontinuum Through a Large Mode Area Photonic Bandgap Fiber

S. Ghosh*, T. Naresh, R. K. Varshney, and B. P. Pal

*Indian Institute of Technology Delhi, New Delhi, India – 110016.*
*Corresponding author: somiit@rediffmail.com*

**Abstract:** We report generation of broadband low divergent supercontinuum over the entire wavelength window of 1.5 to 3.5 μm from a 2.25 meter long effective single moded photonic bandgap fiber with mode area of 1100 μm$^2$.
**OCIS codes:** (060.4370) Nonlinear Optics, Fiber; (190.4380) Nonlinear Optics, Four wave mixing; (190.4410) Nonlinear Optics, Parametric process.

## 1. Introduction

Mid-IR photonics, which has potential for various applications in defense, astronomy, spectroscopy, sensing and biomedical optics, has evolved as an important area of contemporary research [1, 2]. Further, it has also been realized that for better understanding of the underlying fundamental physics in the context of molecular/atomic spectroscopy, and astronomy, the optical test-bed could be exploited. With the advent of reasonably transparent materials like soft glasses, chalcogenide glasses, fluoride glasses in the wavelength range of 2-10 μm and development of mature fabrication technology based on extrusion, and other suitably modified methods like CVD/MCVD, the ongoing research to develop devices and components for mid-IR applications has given further flip in this direction. Optical fiber-based discrete/continuum mid-IR compact light sources are highly desirable as one of the building blocks for these applications. Owing to their unique dispersion and nonlinear characteristics, and high degrees of freedom for parametric control to tailor their optical properties for specialty applications, photonic bandgap guided all-solid Bragg fiber is one of the strong candidates for wavelength translation or new frequency/ supercontinuum (SC) generation [3, 4]. Bragg fiber designs are amenable to large mode effective areas for effective single-mode guidance, which is attractive for high power delivery without nonlinear optical impairments, besides being less prone to bend-loss compared to any of its counterpart microstructured fiber geometries [2]. Moreover, continuum generation using large mode area (LMA) fibers is particularly attractive for achieving high throughput power, which is difficult with small core fibers due to the limitations imposed by the damage threshold of the fiber material [5-7]. Moreover, when SC is generated through highly nonlinear fibers with small core area, beam divergence of the output beam could be large, additional optics is normally required for coupling to another fiber.

In this paper we report design of a soft glass-based Bragg fiber [2] tailored for generation of high power low divergent supercontinuum light extending from 1.5 – 3.5 μm from a 2.25 *m* long fiber when a 1 *ps* seed pulse centered at 2.04 μm was assumed to be launched at its input end. The fiber has been designed with its zero dispersion wavelength (ZDW) as 2.04 μm close to that of the pump laser. The designed fiber would exhibit negligibly small divergence (NA of the fiber ~ 0.035). The chosen soft glass-based material systems (SF6 and LLF1) have good transparency at the proposed wavelength range of operation and are fabrication friendly for realizing fibers through extrusion technique.

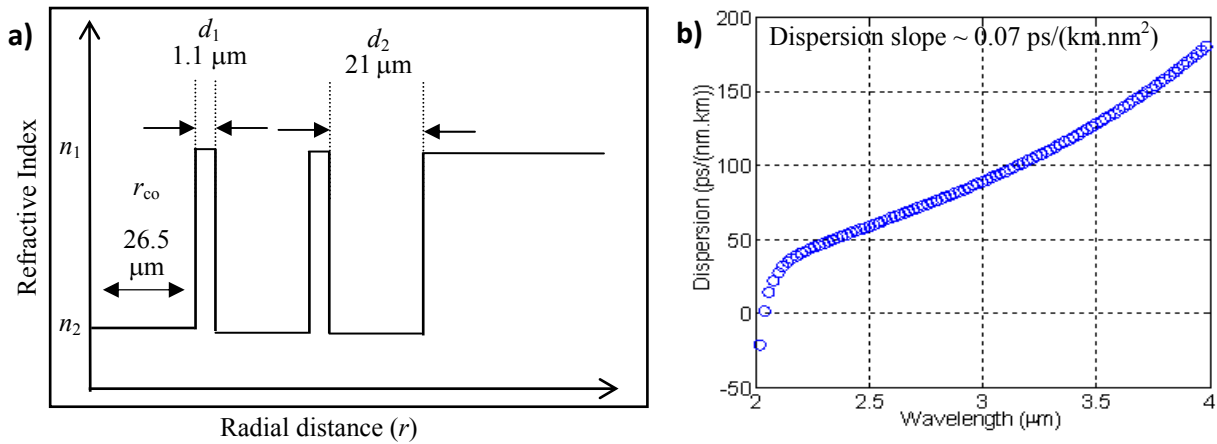

Fig. 1. a) Schematic of refractive index profile of the proposed effective single moded LMA fiber; b) dispersion spectrum of the guided fundamental mode of the LMA fiber.

## 2. Modal analysis and propagation study

To optimize the refractive index profile of the Bragg fiber (as shown in Fig. 1(a)), its modal analysis has been carried out in FEM-based software Comsol®. The effective mode area of the effectively single-mode guiding multimode fiber [2] after a distance of only 2 *m* is estimated to be ~ 1100 μm$^2$. Figure 1 (b) shows the total dispersion coefficient (*D*) vs wavelength of the designed Bragg fiber, wherefrom it is evident that the ZDW of the fiber is 2.04 μm and the dispersion slope is negligibly small (~ 0.07 ps/(km.nm$^2$) @ 3 μm). This particular ZDW is very interesting as this is ideal for optical pumping with commercially available high power pulsed Thulium/ Holmium doped fiber lasers. To study short pulse propagation though the designed LMA fiber, we solve the generalized nonlinear Schrödinger equation (GNLSE). The GNLSE includes effects of higher-order dispersion and stimulated Raman scattering and can be written as [8]

$$\frac{\partial A}{\partial z} + \frac{i}{2}\beta_2 \frac{\partial^2 A}{\partial t^2} - \frac{1}{3!}\beta_3 \frac{\partial^3 A}{\partial t^3} + \ldots + \frac{\alpha}{2}A = i\gamma\left(1 + \frac{i}{\omega_0}\frac{\partial}{\partial t}\right)\left(A(z,t)\int_0^\infty R(t')|A(z,t-t')|^2 dt'\right) \quad (1)$$

where α is the fiber loss and $\beta_i = \frac{\partial^i \beta}{\partial \omega^i}$, where i = 2,3…7 are the higher order dispersion terms. The nonlinear coefficient γ accounts for the intensity dependence of the refractive index. $R(t')$ is the nonlinear response function, which accounts for the Raman contributions, and can be written as

$$R(t') = (1 - f_R)\delta(t - t_e) + f_R h_R(t) \quad (2)$$

where $f_R$ = 0.13 [3] and $h_R(t)$ is the Raman response function, contains information on the vibration of host material (within the optical fiber) as light passes through the fiber and can be expressed as [8]

$$h_R(t) = \frac{\tau_1^2 + \tau_2^2}{\tau_1 \tau_2^2} e^{\left(\frac{-t}{\tau_2}\right)} \sin\left(\frac{-t}{\tau_1}\right) \quad (3)$$

where $\tau_1$ = 5.5 fs and $\tau_2$ = 32 fs [3] for the chosen soft glass system. We solve the nonlinear Schrödinger equation using split step Fourier method mentioned in [8] to simulate supercontinuum generation in the designed LMA.

## 3. Generated supercontinuum and its evolution dynamics

We have numerically studied pulse broadening through the designed LMA fiber over various lengths of propagation for different sets of the pulse parameters and maximized the wavelength translation efficiency along with nonlinear spectral broadening. Accordingly, we have chosen the central operating frequency of the pulse very close to the ZDW of the LMA fiber for achieving efficient four wave mixing process for wavelength translation. For a 1 *ps* FWHM 20 kW secant temporal pulse as input, the generated SC spectrum at the end of a 2.25 *m* long fiber is shown in Fig. 2(a). Figure 2(b) shows the corresponding dynamics for pulse evolution. To understand the governing nonlinear dynamics leading to this SC spectra for this chosen set of pulse and fiber parameters, we have estimated the dispersion length ($L_D$) to be 150.4 *m*, nonlinear length ($L_{NL}$) to be 0.025 *m* and $L'_D$=588.11 *m*. From the estimated values of $L_D$, $L_{NL}$ and $L'_D$, one can appreciate that nonlinear effects dominate over the dispersion (since $L'_D \gg L_D \gg L_{NL} \approx L$) effects [4]. In Fig 2(c), we have shown the pulse spectra at 25 *mm*, 125 *mm* and 650 *mm* lengths of the fiber. This plot evidently carries the signature of pulse broadening due to self phase modulation (SPM). Similarly, to investigate the broadening dynamics, we have plotted the pulse intensity spectra at 0.85 *m*, 0.9 *m* and 0.95 *m* lengths of the fiber. Figure 2(d) clearly manifests the contribution from FWM process near the shorter and central wavelength part and soliton fission process at the longer wavelength side of the spectrum. The appearance of the unwanted peak around 2.5 μm can be attributed to the onset of soliton fission process. To verify the same, we have calculated the fission length (the length over which the soliton fission occurs) ($L_{fiss}$) to be 1.93 *m* [4]. This estimated length scale has a clear correspondence with the results depicted in Fig. 2(b) as it can be clearly observed that the soliton fission exactly occurs at the same length. Hence, the dip in the SC broadening is due to the breaking occurring as a result of soliton generation at that wavelength. We may mention that we have neglected the effect of loss of the fiber within this wavelength window while studying pulse propagation.

To conclude, in this paper we propose generation of high power low divergent mid-iR SC spanning over the wavelength range of 1.5 to 3.5 μm. An all solid, soft glass-based effective single-mode guiding multimode Bragg fiber with large mode area of length 2.25 *m* has been used for this study. This compact, high power SC source with negligible divergent output, if realized in practice, should be primarily useful for mid-IR optical coherence tomography and spectroscopy.

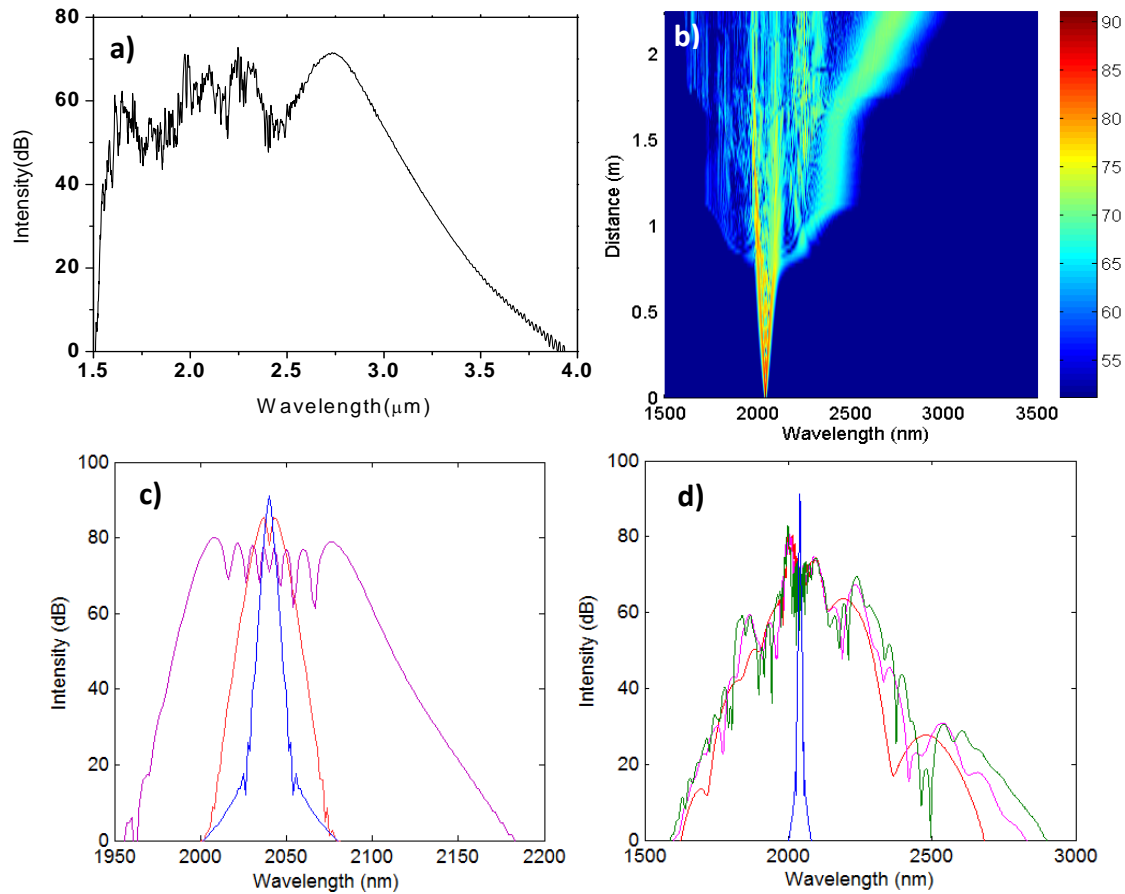

Fig.2: a) Simulated supercontinuum spectra from a 2.25 *m* length of the designed LMA fiber; b) evolution of the short pulse propagation along the fiber length; c) pulse broadening due to SPM during the initial stage at three different lengths of 25 *mm* (blue), 125 *mm* and 650 *mm*, respectively) of pulse propagation; d) pulse spectra to visualize the broadening of the pulse mainly due to solitons fission at 0.85 *m* (red), 0.9 *m* and 0.95 *m* (green).

This work relates to Department of the Navy Grant N62909-10-1-7141 issued by Office of Naval Research Global. The United States Government has royalty-free license throughout the world in all copyrightable material contained herein.